\documentclass[preprint]{aastex}

\shorttitle{Coronal Rain Dynamics}
\shortauthors{Fang et al.}

\begin{document}

\title{Multidimensional modeling of coronal rain dynamics}
\author{X.~Fang, C.~Xia, and R.~Keppens}
\affil{Centre for mathematical Plasma Astrophysics, Department of Mathematics, 
KU Leuven, Belgium}

\begin{abstract}
We present the first multidimensional, magnetohydrodynamic simulations which 
capture the initial formation and the long-term sustainment of the enigmatic 
coronal rain phenomenon. We demonstrate how thermal instability 
can induce a spectacular display of in-situ forming blob-like 
condensations which then start their intimate ballet on top of 
initially linear force-free arcades. Our 
magnetic arcades host chromospheric, transition region, and coronal plasma. 
Following coronal rain dynamics for over 80 minutes physical time, we collect
enough statistics to quantify blob widths, lengths, velocity distributions, and
other characteristics which directly match with modern observational knowledge.
Our virtual coronal rain displays the deformation of blobs into 
$V$-shaped like features, interactions of blobs due to mostly pressure-mediated 
levitations, and gives the first views on blobs which evaporate in situ, or get 
siphoned over the apex of the background arcade. Our simulations pave the way 
for systematic surveys of coronal rain showers in true multidimensional settings, 
to connect parametrized heating prescriptions with rain statistics, ultimately
allowing to quantify the coronal heating input.
\end{abstract}


\section{Introduction}
A recurrent finding in coronal loops is the coronal rain phenomenon, seen as
intensity variations signaling cool blob-like downflows along the legs of loops 
\citep{kawaguchi70,leroy72,schrijver01,oshea07}. 
Coronal rain forms part of the general phenomenon of thermal instability in a 
plasma, that takes place whenever radiative losses locally overcome the heating 
input \citep{parker53,field65}, and is related to ``catastrophic cooling" events
\citep{schrijver01}. Meanwhile, numerical studies have significantly 
contributed to the understanding of these events, but typically adopted 
simplifying one-dimensional (1D) approximations meant to demonstrate the 
thermodynamic evolution along individual field lines \citep{goldsmith71, mok90, 
antiochos91, antiochos99, chun11}. For coronal rain to occur in loops, the 
heating input is generally accepted to be concentrated at the loop footpoints. 
With footpoint heating, the loops rapidly get hotter and denser, due to 
evaporated chromospheric plasma invading the loops. The combined action of 
anisotropic thermal conduction and optically thin radiation causes these coronal
hot loops to ultimately reach thermally unstable regimes in a timescale of hours. 
After that, ``catastrophic cooling" sets in locally, leading to the rapid 
formation of condensations, as demonstrated in 1D models \citep{karpen01,
muller03,muller04,muller05,groof05,antolin10,chun11}. In this paper, we present 
the first numerical study of the coronal rain phenomenon in a 2.5-dimensional 
model where a magnetic arcade hosting chromospheric, transition region, and 
coronal plasma demonstrates a coronal rain shower lasting for over an hour. 
This allows us to collect statistical information that can
confront recent observational insights. 

From the observational side, the various stages of coronal rain formation have 
been analysed using \emph{TRACE}, and were found to be recurring in timescales 
of days to weeks \citep{schrijver01}. 
Observations of coronal rain with \emph{Hinode}/SOT have revealed a clear 
thread-like character in the coronal loops, and have started to provide 
statistical info on the number and velocities of blobs, while sizes reach down 
to the resolution limits \citep{antolin10}. High resolution
instruments now reveal
a scenario that coronal rain is a rather common phenomenon \citep{kamio11,
antolin11,antolin12}, and can provide key info on the elusive coronal heating 
problem itself~\citep{antolin10}. Realizing multi-dimensional numerical studies 
will be a prerequisite to unravel how coronal rain statistics encodes this 
heating input.

\section{Numerical setup}\label{setup}
Our simulation uses a 2.5D thermodynamic MHD model as in \citet{chun2012}, 
on a 2D domain of size 80 by 50 Mm (in $x-y$). The initial magnetic topology now adopts a linear 
force-free magnetic field characterized by a constant angle $\theta_0$ as 
follows:
\begin{displaymath}
 B_{x}=-B_{0} \cos \left( \frac{\pi x}{L_{0}} \right) \sin\theta_0 \exp\left(
 -\frac{\pi y \sin\theta_0}{L_{0}} \right)\,,
\end{displaymath}
\begin{displaymath}
 B_{y}=B_{0} \sin \left( \frac{\pi x}{L_{0}} \right) \exp\left(
 -\frac{\pi y \sin\theta_0}{L_{0}} \right)\,,
\end{displaymath}
\begin{equation}
 B_{z}=-B_{0} \cos \left( \frac{\pi x}{L_{0}} \right) \cos\theta_0 \exp\left(
 -\frac{\pi y \sin\theta_0}{L_{0}} \right)\,.
\end{equation}
Setting $\theta_0=30^\circ$ corresponds to the arcade making a $30^\circ$ angle 
with the neutral line. $L_{0}=80$ Mm is the horizontal size of our domain, and 
adopting $B_{0}=12$ G leads to a realistic 2.5D magnetic topology.

To obtain a self-consistent thermally structured corona, we augment 
this setup with a background heating rate decaying exponentially with height,
\begin{equation}
 H_{0}=c_{0} \exp\left(-\frac{\sqrt{2} \pi y }{2L_{0}} \right) 
 \,,
\end{equation}
where $c_{0}=10^{-4}$ erg cm$^{-3}$ s$^{-1}$. This initial setup is out of 
thermal equilibrium, so we need to integrate the governing equations in time 
with $H=H_{0}$ active until the system relaxes to a 
quasi-equilibrium state. 

We use the parallelized Adaptive Mesh Refinement (AMR) Versatile Advection Code 
\citep{amrvac}. Our domain has an initially symmetric setting in area $-40 <  x < 40$ Mm 
and $0 <  y < 50$ Mm. An effective resolution of $1024 \times 640$ is attained 
by using four AMR levels, with an equivalent spatial resolution of 78 km in 
both directions. 

Using this numerical strategy, the configuration reaches a quasi-equilibrium 
state. An online movie shows the temperature evolution through this relaxation phase, which already demonstrates some thermodynamic
structuring in the final arcade. Flows are forced into standing wavelike patterns in the 2D arcade, 
and the overall evolution gradually damps their kinetic energy, so relaxation is identified as the time when the maximal residual velocity in the 
domain has become less than 5 km s$^{-1}$. In that end state, a comparatively 
thin transition region connects chromosphere to corona, and is located 
at heights between 3 Mm and 5 Mm. The plasma beta is 0.07 at 20 Mm 
height above the neutral line while the temperature and number density there are
individually around 1.7 MK and $3.3 \times 10^8 {\mathrm{cm}}^{-3}$. Beginning 
with this equilibrated system, we add a relatively strong heating
$H_{1}$. This extra heating is localized near the chromosphere with the formula
as \citep{chun2012}:
\begin{equation}
 H_{1}=\left\{
\begin{array}{lrrrr}
C_{1}  & {\mathrm{if}} & { y<y_{c}} & {\mathrm{and}} & 
{A_{1}(2.6)<A(x,y)<A_{1}(1.4)}\\
C_{1} \exp(-(y-y_{c})^{2}/\lambda^2) & {\mathrm{if}} & {y\geq y_{c}} & 
{\mathrm{and}} & {A_{1}(2.6)<A(x,y)<A_{1}(1.4)}
\end{array} \right.
\end{equation}
\begin{displaymath}
A_{1}(x)=\frac{B_{0}L_{0}}{\pi}\cos\left(\frac{\pi x}{L_{0}}\right) \,,
A(x,y)=A_{1}(x)\exp\left(-\frac{\pi y \sin\theta_{0} }{L_{0}} \right) \,,
\end{displaymath}
\begin{displaymath}
\lambda^2=\frac{8\left(A(x,y)-A_{1}(1.4)\right)}{A_{1}(2.6)-A_{1}(1.4)}+12
\rm\quad (Mm^{2}) \,,
\end{displaymath}
where $C_{1}=10^{-2}$ erg cm$^{-3}$ s$^{-1}$, $y_{c}$=3 Mm and $\theta_0=
30^\circ$. This choice of strong base heating contrast ($C_{1}/c_{0}=100$),
can mimick extra heating provided by a flaring event, and helps to reach the relevant dynamical phase at earlier times.

\section{CORONAL RAIN FORMATION AND STATISTICS}

\begin{figure}
 \centering
 \includegraphics[width=\textwidth]{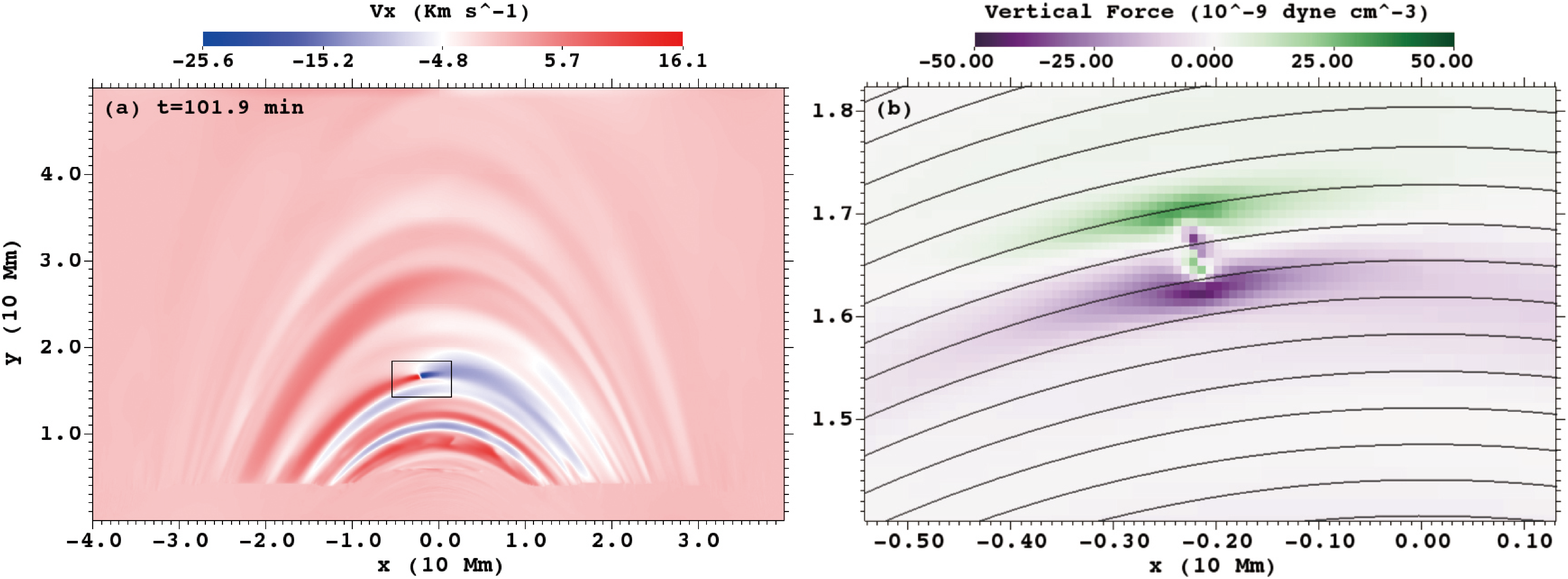}
 \caption{At $t\approx 100$ minutes, we show the $x$-velocity component at left. 
 Right panel: zoomed view on the local signed 
  vertical total force.}
 \label{velocity}
\end{figure}

Because the heating 
formula for $H_1$ affects only a selection of loops fully contained interior to 
our simulation domain, this part of the arcade witnesses increased 
densities and temperatures, with maximum values of 2.1 MK after 9 minutes of added heating. 
Despite loop-aligned thermal conduction transporting energy to the dense coronal
plasma around the apexes, temperatures then start to reduce slowly,
while the densities still keep increasing. The locally heated arcade system 
continues to evolve, and only after about 100 minutes of sustained heating, the 
temperature at a height of 16.5 Mm suddenly declines drastically to 0.04 MK, 
slightly off-center. A small condensation segment with a density 
$5.6 \times 10^{10}$ cm$^{-3}$ suddenly comes forth around the apexes of a 
strand of magnetic loops. Figure~\ref{velocity} shows the velocity field, and the (signed) vertical total
 force with gravity, Lorentz force and pressure gradient in a zoomed view on the blob forming. The overall perturbed 
force field extends over 1 Mm in width, and has dominant about equal and in-phase pressure and Lorentz force contributions 
and induces field variations on neighboring fieldlines, which aid in triggering sympathetic condensations. Indeed, after this first localized condensation event, similar
condensation processes continuously arise on both ends of the first 
condensation. Due to the broken symmetry, we observe this to extend into coronal
loops on either side of the first affected loop strand, and this results into 
the larger scale condensation to look like a zigzag rope (like in panel (c) of 
Fig.~\ref{process}). What happens next is a spectacular display of fragmenting, 
forming, relocating plasma blobs, since the cool plasma condensations 
spontaneously loose their balance between existing forces (gravity, magnetic, 
and gas pressure gradients), and start to slide down slowly along magnetic 
field lines. In the online movies, one can see how at about 118 minutes, the big zigzag condensation begins to 
split into several smaller blobs, descending along both rims of the magnetic field.
After about 160 minutes, also due to the depletion of plasma in these loops, 
the subsequent phase seems less vigorous. Similar phases 
can be found in observations \citep{antolin10,antolin12}, 
and are interpreted as `limit cycles of loop evolution' by \citet{muller03}.
   
Our simulation shows new features related to blob destruction. In particular, 
at 167 minutes, at a height of 10 Mm and horizontal position of $x=-21$ Mm, a 
small baby blob with a number density of $9.0 \times 10^{9}$ cm$^{-3}$ and 
temperature 0.55 MK, forms in a first slowly upflowing part of a strand of 
loops, where another bigger blob has just descended. This blob has an upward 
velocity of 10 km s$^{-1}$, but then gets destroyed by a hot inflow from the 
other side due to the heating-induced evaporation at the other loop footpoint. 
 This is supported by an online movie.

At the overall effective resolution, any individual grid cell where the number 
density exceeds $7.0 \times 10^9 {\mathrm{cm}}^{-3}$, the temperature drops 
below 0.1 MK in the corona is labeled as in a coronal rain blob. These
threshold values are suggested by observational findings \citep{hirayama85} and 
other numerical simulations \citep{muller05,antolin10}. To count the 
instantaneous amount of blobs present at one time, we then identify the total 
number of blobs by assuming that all connected labeled pixels actually 
compose a single blob. In that way, we can report on the instantaneous amount of
coronal rain blobs and the centroid $(x_c,y_c)$ coordinates of each blob. 
The local magnetic field vector defines directions along and 
perpendicular to the field line. Along these directions, the length and width of
the blob are quantified. However, since the resolution of our numerical 
simulation (78 km) is much higher than current observational resolutions, e.g., 
150 km of CRISP \citep{antolin12}, the number of identified blobs in the 
simulation is larger than that found in comparable observations. For the sake 
of direct comparison with the observations, we also do this at a
resolution of 200 km. This operation combines neighboring blobs and occasionally 
overlooks blobs with sizes below this resolution.

Fig.~\ref{time}(a) shows that the total mass of all blobs as function of time is 
nearly identical between the numerical resolution (dashed curve) and the 
observational resolution (solid curve), while the former slightly exceeds the 
latter. The difference between observations versus simulations is more 
pronounced in Fig.~\ref{time}(b) showing the actual numbers of blobs. While 
actual blob numbers can go over 100 at certain times, still when viewing them 
with observational resolution as fewer (less than 20) blobs, the total mass 
basically remains the same between different resolutions. This means that while 
current coronal rain related mass estimates from observations are likely to be 
correct, there are still a great quantity of small unresolved blobs in 
present-day observations. After the first condensation seen at $t\approx 100$ 
minutes, Fig.~\ref{time}(a) shows that in the next 29 minutes, still before the 
first descending blob crashes into the transition region, the mass accumulation 
of the blobs scales at a rapid rate of 6.7 g cm$^{-1}$ s$^{-1}$. To quantify a 
true mass drain rate, we could adopt an average size in $z$ of 400 km 
as the average width, making the mass drain rate about $3\times10^{9}$ 
g s$^{-1}$, very similar to observational results \citep{antolin12}. 
Snapshots of density and temperature at times $t\approx 136$ and $t\approx 152$ 
minutes are shown in Fig.~\ref{process}, where selected blobs are labeled by 
numbers, used in the further discussion.

\begin{figure}
 \centering
 \includegraphics[width=6cm,angle=-90]{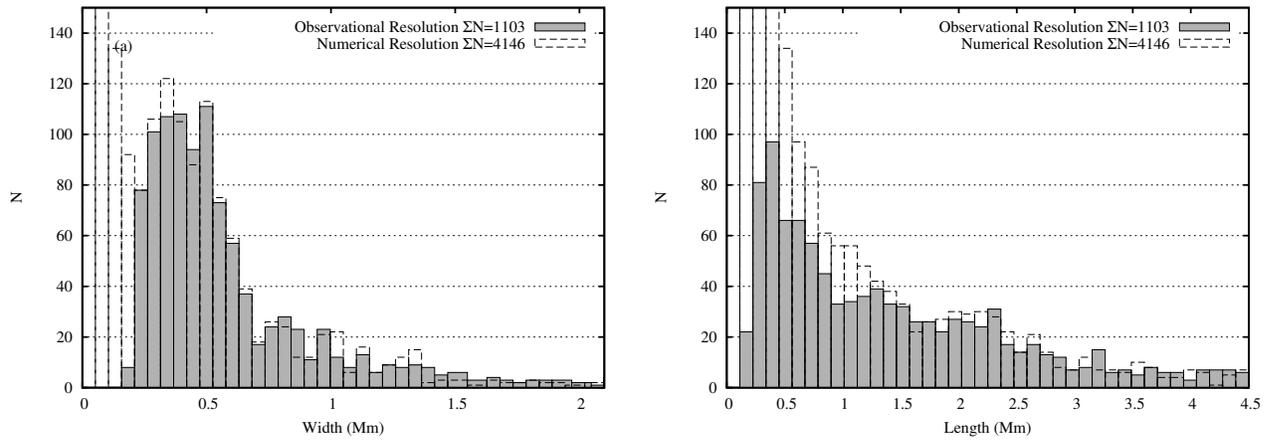}
 \caption{(a) and (b) show the distribution function of width and length, 
 respectively, at numerical and 
 observational resolution.}
 \label{function}
\end{figure}

A large variety of blob appearances are found during the whole coronal rain 
process. By treating every snapshot between $t\approx 100$ and $t\approx 200$ at
a time interval of 43 seconds as an individual observation, we can easily 
obtain statistically meaningful distribution functions of blob width and length.
This is quantified in Fig.~\ref{function} where we again contrast findings based on 
the numerical resolution with the observational resolution.
The width of the blobs reveal the intrinsic cross section of a strand of loops 
with nearly synchronous evolution. Recent results from triple-filter analysis of
the finest coronal loops analyzed in TRACE images found elementary loop strands
with isothermal cross sections of $\approx  1000-2000$ km \citep{aschwanden05}.
Similar values of sympathetic loop strand widths can be seen in the horizontal 
velocity map in Fig.~\ref{velocity}, are also seen in the perturbed force view and return in the distribution function of the obtained 
blob widths in panel (a) of Fig.~\ref{function}. Although the width of such strands 
and blobs can reach the maximum value of 2000 km, these huge blobs will be 
separated during the propagation process into small fragments. This is again 
resulting from significant differences in the diverse forces acting along their 
body. The width histogram in Fig.~\ref{function}(a) also shows that the vast majority
of blobs possess widths like $\approx 200-1000$ km with an average 400 km, in 
direct correspondence with recent observational results from \cite{antolin12}. 

\section{DYNAMICS OF BLOBS}

\begin{figure}
\centering
\includegraphics[width=\textwidth]{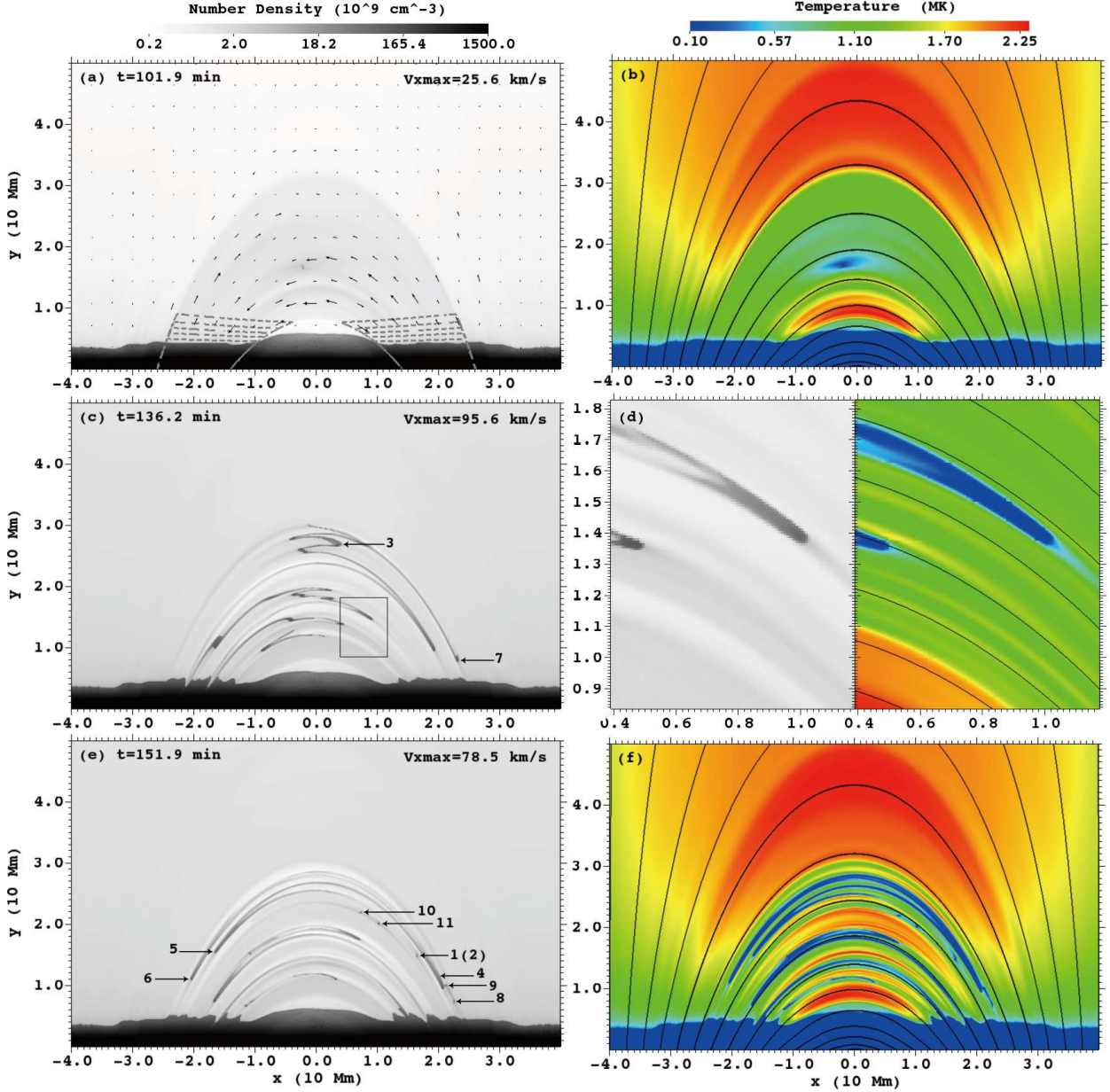}
\caption{Snapshots at t=100 (first row), 136 (second row) and 152 minutes 
(third row). At left: density. At right: temperature and magnetic field lines.
Localized heating is shown as contours in panel (a) and velocity arrows. 
 Panel (d) shows the thermal stucture, zooming into blobs in panel (c).}
\label{process}
\end{figure}

The velocity structure at $t\approx 100$ minutes is shown in Fig.~\ref{velocity}, 
at the same time as the density and temperature panels (a) and (b) of 
Fig.~\ref{process}. In the velocity plot, one identifies the condensation where 
two strong opposite inflows with a maximum relative velocity of 68.7 km 
s$^{-1}$ are siphoned towards the condensation site from both sides.
This coincides with a dramatic evacuation of a loop strand caused by the 
catastrophic cooling. The thermodynamic evolution rapidly refills the local 
empty loops with hot and rarefied plasma. These fast inflows and the density 
variation they create, first realize a pressure difference across the two sides 
of the off-center blob, which levitates the newborn blob against gravity. This 
first phase impedes the descending process of newborn blobs. However, after a 
short time the inflows become slower, and while the blob density increases, this
 previous pressure difference gradually fades away. Therefore, they ultimately 
start to accelerate quickly downwards. 

To explain how a full loop strand ultimately shows blobs that appear like
comet-shaped or $V$-like features during propagation \citep{antolin10,
antolin11}, we note that within a loop strand of finite extent (say few hundred 
km in width), a first small condensation functions like the seed for a larger 
blob. In this growth, the condensation process appears to extend from the first 
blob onwards due to the synchronous temperature evolution in a wider loop strand
\citep{klimchuk10}. This means that while the firstly formed condensation may 
already have evolved beyond the phase where it experiences levitating pressure 
support, the condensation segments formed later at the edge of the blob are still locally
supported against gravity by the pressure difference due to the fast siphon 
inflows. As a result, the large, growing blob gets deformed as a whole into a 
comet-shaped pattern, like the blobs labeled with numbers 3-7 of Fig.~\ref{process}.
During their propagation towards the arcade footpoints, catastrophic cooling 
further sets in in the tail of these blobs, and blobs will be elongated by 
continuously forming condensations on the way down. Furthermore, as the 
gravitational acceleration varies with height, an effect accounted for in our 
external $y$-stratified gravitational field, the blob will also become elongated
due to being stretched by the differential component of gravity along the curved magnetic field. 
Therefore, the length histogram in panel (b) of Fig.~\ref{function}
presents an average of 850 km for coronal rain blobs, but shows a wide range of
lengths going from 200 km to exceeding 4500 km, a fact confirmed by 
observations \citep{schrijver01,antolin12}.
Zoomed views on selected blobs in Fig.~\ref{process} show the local temperature structure, with
conduction-dominated regions around the blobs. The temperatures of these local transition regions are around 0.6 
MK.

\begin{figure}
 \centering
  \includegraphics[width=12cm,angle=-90]{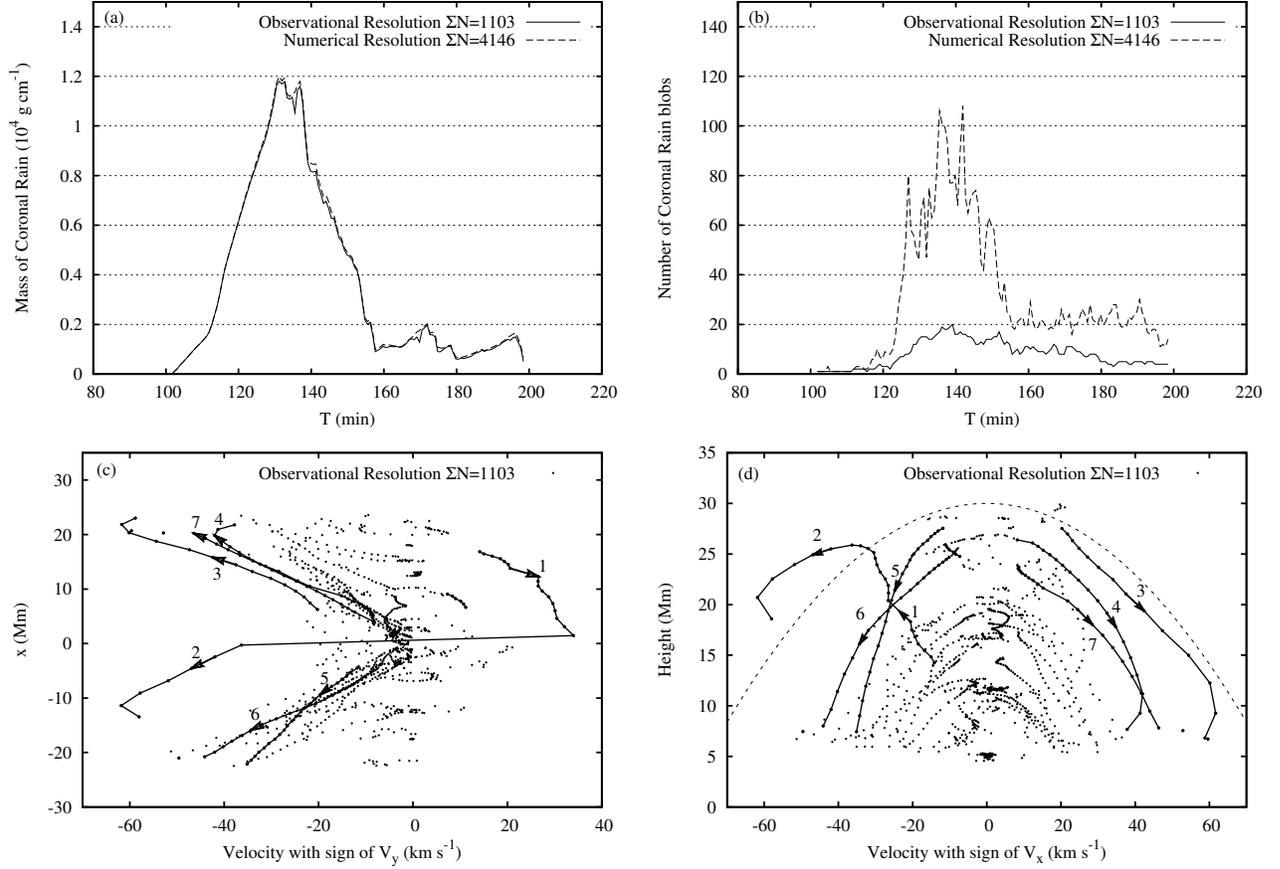}
 \caption{(a)Total mass versus time with numerical resolution (dashed curve) 
and observational resolution (solid curve); (b) number of blobs; (c) scatter 
plot of projected velocity with sign of $v_{y}$ versus $x$-axis value; (d)
scatter plot of projected velocity with sign of $v_{x}$ versus height of blobs. 
The dashed curve denotes the path that a blob would follow if falling from a 
height of 30 Mm and subject to an acceleration of 0.18 km s$^{-2}$. All solid 
curves connecting points in (c) and (d) show the trace of several blobs, 
numbered from 1(2) to 7.}\label{time}
\end{figure}

We obtain a broad distribution of projected velocities, ranging from few km s$^{-1}$ 
to the high velocity of descending blobs going up to more than 60 
km s$^{-1}$. Panel (c) of Fig.~\ref{time} shows a scatter plot of the horizontal 
centroid $x_c$-position of the blobs versus their in-plane projected velocity, 
signed by vertical velocity. This is done at the observational resolution, and 
in this view one can trace individual blobs appearing in multiple snapshots. 
Panel (d) of Fig.~\ref{time} shows a scatter plot of height $y_c$ of the blobs 
versus their projected velocity, now signed with $v_{x}$. 
Since the velocities are generally height dependent, the dashed curve 
in panel (d) of Fig.~\ref{time} denotes the path that a blob would follow if it 
were falling from a height of 30 Mm, subject to an acceleration of 0.18 km 
s$^{-2}$, the average effective gravity for a loop whose height to half
baseline ratio is 30 (Mm)/26 (Mm). We note that most of the measurements are
located below the dashed curve, like those for blob 6 and 7. This
scenario suggests a role for other forces than gravity, like gas 
pressure as suggested by previous 1D numerical simulations \citep{muller03,
muller04,muller05}.

Close to the lower parts above the transition region of the arcade, strong deceleration of individual 
blobs are sometimes observed \citep{antolin12}, which is explained by the increase of gas 
pressure there from the higher local densities. The solid lines connecting the 
points of individual blobs 3 and 4 in panels (c) and (d) of Fig.~\ref{time} show 
these strong decelerations happening right above the transition region. 
Decelerated by this pressure gradient, the leading descending blob part
could be caught up by a later faster descending blob part (as in 1D studies 
from \citep{muller05}) and merge to one heavier blob. At about 152 minutes, in 
panel (e) of Fig.~\ref{process}, at a height of 7.1 Mm and horizontal position of 
$x=22.5$ Mm, we find that in the trail of a formerly descending blob, a small 
blob (number 8) appears and stays there supported by the large 
pressure gradient. Meanwhile, in the same strand, another blob (number 9) 
forms above the number 8 and moves towards it with velocity of 26 km s$^{-1}$. 
They collide, merge and produce a heavier blob, which finally falls down 
to the transition region 4 minutes later. A movie with a zoomed view on this process is available online.

In panel (e) of Fig.~\ref{process}, two blobs in the same flux loop strand, numbered
10 and 11, approach each other because of the significant pressure difference 
across them, as extremely low gas pressure is induced by catastrophic cooling in
between them, and the gas pressure outside enforces their mutual approach. This
kind of situation can even suck a blob upwards, ascending and crossing the 
apexes of loops, e.g. this is what happens to the blob number 1(2) in panel (e) 
of Fig.~\ref{process}, which shares the same strand with blob number 5. In panels 
(c) and (d) of Fig.~\ref{time}, the scatter velocity plots versus height and 
$x$-position show us this clearly when inspecting the traces of blob 1(2) and 
blob 5. When blob 5 descends to the footpoint, blob 1(2) is siphoned to ascend 
from the right rim and over to the left rim along the magnetic field lines. 

\section{Conclusions}
We simulate the initial formation and the long-term sustainment 
of the enigmatic coronal rain phenomenon for the first time in a realistic 
multi-dimensional magnetic configuration. In the over 80 minutes physical time, 
we collect enough statistics to quantify blob widths, lengths which average 
400 km, 800 km, and the velocity distribution from small values to 65 km s$^{-1}$. 
Our virtual coronal rain display features the deformation of blobs into 
$V$-shapes, interactions of blobs due to mostly pressure-mediated 
levitations, and gives the first views on blobs which evaporate in situ, or get 
siphoned over the apex of the background arcade.
We will perform parameter studies for similar arcade configurations, varying 
field strength, overall topology and the role of magnetic shear. 

\acknowledgments
Results obtained in the projects GOA/2009/009 (KU Leuven), the EC Seventh 
Framework Programme (FP7/2007-2013) under grant agreement SWIFF (project no. 
263340, {\tt www.swiff.eu}), and by the Interuniversity Attraction Poles 
Programme initiated by the Belgian Science Policy Office (IAP P7/08 CHARM). Part
 of the simulations used the infrastructure of the VSC - Flemish Supercomputer 
Center, funded by the Hercules Foundation and the Flemish Government - 
Department EWI. We acknowledge fruitful discussions with P. Antolin and T. Van 
Doorsselaere, and helpful comments from a referee.

\normalfont

\end{document}